# Multivariate Analysis on Performance Gaps of Artificial Intelligence Models in Screening Mammography

**Linglin Zhang, MS[1], Beatrice Brown-Mulry, BS[1], Vineela Nalla, MS[2], InChan Hwang, MS[1], Judy Wawira Gichoya, MD[3], Aimilia Gastounioti, PhD[4], Imon Banerjee, PhD[5,6], Laleh Seyyed-Kalantari, PhD[7], MinJae Woo, PhD[1], Hari Trivedi, MD[3]**

1. School of Data Science and Analytics, Kennesaw State University, 3391 Town Point Dr NW, Kennesaw, GA 30144
2. Department of Information Technology, Kennesaw State University, 1100 South Marietta Pkwy, Marietta, GA 30060
3. Department of Radiology and Imaging Sciences, Emory University, 1364 E Clifton Rd NE, Atlanta, GA 30322
4. Computational Imaging Research Center, Washington University in St. Louis School of Medicine, 4525 Scott Avenue, St. Louis, MO 63110
5. Department of Radiology, Mayo Clinic Arizona, 13400 E Shea Blvd, Scottsdale, AZ 85259
6. School of Computing and Augmented Intelligence, Arizona State University, 699 S Mill Ave, Tempe, AZ 85281
7. Department of Electrical Engineering and Computer Science, York University, 4700 Keele St, Toronto, Ontario, Canada M3J 1P3

{lzhang23, bbrownmu, vnalla, ihwang}@students.kennesaw.edu
{judywawira, hari.trivedi}@emory.edu
a.gastounioti@wustl.edu
banerjee.imon@mayo.edu
lsk@yorku.ca
mwoo1@kennesaw.edu

**Abstract**

Although deep learning models for abnormality classification can perform well in screening mammography, the demographic, imaging, and clinical characteristics associated with increased risk of model failure remain unclear. This retrospective study uses the Emory BrEast Imaging Dataset (EMBED) containing mammograms from 115,931 patients imaged at Emory Healthcare between 2013-2020, with Breast Imaging Reporting and Data System (BI-RADS) assessment, region of interest coordinates for abnormalities, imaging features, pathologic outcomes, and patient demographics. Multiple deep learning models were trained to distinguish between abnormal tissue patches and randomly selected normal tissue patches from screening mammograms. We assessed model performance by subgroups defined by age, race, pathologic outcome, tissue density, and imaging characteristics and investigated their associations with false negatives and false positives. We also performed multivariate logistic regression to control for confounding between subgroups. The top-performing model, ResNet152V2, achieved accuracy of 92.6% (95%CI=92.0-93.2%), and area under the receiver operating characteristics curve 0.975 (95%CI=0.972-0.978). Before controlling for confounding, nearly all subgroups showed statistically significant differences in model performance. However, after controlling for confounding, we found lower false negative risk associates with Other race (RR=0.828;p=.050), biopsy-proven benign lesions (RR=0.927;p=.011), and mass (RR=0.921;p=.010) or asymmetry (RR=0.854;p=.040); higher false negative risk associates with architectural distortion (RR=1.037;p<.001). Higher false positive risk associates with BI-RADS density C (RR=1.891;p<.001) and D (RR=2.486;p<.001). Our results demonstrate subgroup analysis is important in mammogram classifier performance evaluation, and controlling for confounding between subgroups elucidates the true associations between variables and model failure. These results can help guide developing future breast cancer detection models.

## Introduction

Breast cancer is the most common cancer in women and causes 42,000 deaths each year in the United States [1]. Early detection via screening mammography has been shown to reduce morbidity and mortality of breast cancers by 38–48% through the recognition of abnormalities such as masses, calcification, asymmetries, and architectural distortions (ADs) [2,3,4]. However, there is significant resource usage for detecting a relatively small number of cancers - for every 1,000 women screened, approximately 100 are recalled, and 30-40 are biopsied for the detection of only 5 cancers [5]. This is largely due to the similar imaging appearance of many benign and malignant findings which necessitates repeat imaging and/or biopsy for definitive diagnosis. Improving differentiation of benign and malignant findings on screening mammography has potential to reduce wastage and decrease patient morbidity [3,6,7].

Computer-aided detection and diagnosis (CAD) systems, including systems for computer-aided detection (CADe) and computer-aided diagnosis (CADx), have formed a major research area to address the challenges in the field of mammographic imaging [8,9,10,11,12,13,14]. Recently, many deep learning models have been developed to assist radiologists in interpreting mammograms with many results showing systems that meet or exceed radiologist performance, however most models focus on whole image classification [15,16,17,18,19,20,21,22]. Object detection of abnormalities in mammography is more challenging and can be described as two steps: (1) detection of a potential abnormal region of interest (ROI); (2) classification of abnormalities according to image characteristics or likelihood of malignancy [6,17,18,19,20,21,22,23,24,25,26,27]. Locating abnormalities in mammography remains a challenging task due to the subtle nature of their appearance, and there is no object detection model that has been both internally and externally validated for localizing findings in breast imaging [28,29,30,31]. This creates a potential gap in knowledge wherein object detection models for breast cancer may underperform on certain subsets of exams.

Failure analysis in medical imaging involves analyzing model outputs to explore root causes of failures and addressing underlying issues [32]. The benefits of such analysis include providing guidance into the model training process and identifying vulnerable populations who may be disadvantaged by the model if the failures are not properly addressed. To-date, the performance of mammogram abnormality detection models has been described only broadly, lacking specific information about underlying characteristics that may lead to failures, such as demographic, imaging, or pathologic factors. This limitation can largely be attributed to the scarcity of granular datasets that incorporate these features [33, 34]. This leads to a knowledge gap and potential unintended harm from the underperformance of AI models in certain demographic subgroups (e.g., race and age), cancer subtypes, or imaging findings [35,36,37,38,39,40,41]. Analyzing these underlying variables of failure is also crucial for uncovering opportunities to enhance the effectiveness and equity of computer-assisted abnormality detection across diverse patient groups and clinical scenarios.

Traditional subgroup analysis of deep learning model performance in medical imaging included using DeLong test, $\chi^2$ analysis, Student's *t*-test, permutation test, and other non-parametric tests on bootstrapped result of receiver operating characteristics curve (AUC) by considering each subgroup independently in a univariate setting [42,43,44,45,46,47]. While providing useful insights on disparity of performance, the potential confounding effects between multiple subgroups has not been properly addressed in most studies.

In this paper, we present a novel method of failure analysis in patch classification for breast cancer screening, identifying the underlying imaging and demographic characteristics that induce model failures, and explore the significance of these features before and after controlling for confounding effect. We train a deep learning model for classification of normal versus abnormal patches of tissue on screening mammography and present a post-hoc analysis to identify imaging, pathologic, and demographic characteristics that result in model underperformance and could interfere with abnormality detection in screening mammography. False positives (type I error) and false negatives (type II error) pose different challenges in mammography abnormality classification, the root causes of which are explained by our method.

## Materials and Methods

The ability to accurately classify abnormalities in patches of mammograms is likely to affect the performance of object detection models for screening mammography. Therefore, we hypothesized that there are subgroups of images for which a classifier will underperform in abnormality detection based on patient demographics, imaging, or pathologic features. To test this, we develop a two-step experiment: (1) training multiple comparative deep learning models for abnormality classification in mammography, and (2) statistically analyzing the correct and incorrect prediction in abnormal and normal tissue patches before and after controlling for confounding between features using multivariate logistic regression (Fig. 1).

**Data Preparation**

In this retrospective study, we used the EMory BrEast imaging Dataset (EMBED) [48] which contains clinical and imaging data for 3.4M mammogram images from 383,379 screening and diagnostic exams of 115,931 patients from 2013 to 2020 acquired at four hospitals at a single academic institution. EMBED includes demographics, imaging features, pathologic outcomes, and Breast Imaging Reporting and Data System (BI-RADS) [49] tissue densities and scores for screening and diagnostic exams. ROIs annotated by the radiologist during original interpretation are also available, predominantly for abnormal screening (BI-RADS 0) exams.

An overview of the data preparation is shown in Fig. 2. For this study, we used only full field digital mammograms (FFDM). Positive patches were defined as ROIs annotated on BIRADS 0 images and labeled with imaging features (mass, calcification, AD, asymmetry) and pathologic outcome (never biopsied, benign, or malignant). The distribution and location of positive patches was calculated and used to inform selection of negative patches in a matching distribution as follows. Negative image patches were generated from two sources: (1) random patches from BI-RADS 0 images while avoiding any annotated ROIs; (2) random patches from negative screening images (BI-RADS 1 and 2) as shown in Fig. 3. All negative patches were selected from regions with breast tissue while avoiding background air by only allowing cropping pixels with <10% zero values. This was done to include tissue that contains the edge of the breast as was observed in positive patches. Patch sizes ranged from 53×76 to 2379×2940 with median size of 360×437. Patches greater than 512×512 were downsampled to less than 512×512 while maintaining aspect ratio. All patches were then zero-padded to be 512×512 pixels for consistent input size for the convolutional neural network (CNN) architecture. The final dataset was randomly divided at the patient level to prevent leakage, and consisted of 10,678 patients and 29,144 patches (55.6%) for training; 3,609 patients and 9,910 patches (18.9%) for validation; and 5,404 patients and 13,390 patches (25.5%) for testing.

**Patch Classifier Model Training**

Guided by previous studies, we tested four common CNN architectures to classify between normal and abnormal tissue patches - InceptionV3 [50], VGG16 [51], ResNet50V2 [52], and ResNet152V2 [53] using the same training, validation, and test sets. Preliminary results demonstrated top performance by ResNet152V2 (Table 1), so this architecture was utilized throughout the study. Input layers were adjusted to accept 512×512-pixel resolution PNG files, followed by ResNet152V2 classifier pretrained on the ImageNet dataset [54], with the additional layers including a trainable batch normalization layer. The following deep neural network constituents included a dense layer with ReLU activation [55], a dropout layer, three dense layers with ReLU, and final dense layer activated by Sigmoid function [56]. Learning rate was optimized using Bayesian optimization [57], with the optimal model validation performance achieved at learning rate = 0.00588461.

To obtain 95% confidence intervals, we performed bootstrapping using two hundred iterations of the whole test set (sampled n ∈, and two hundred iterations in each BI-RADS tissue density group (sampled n ∈ [500, 1666], density A; n ∈ [500, 4921], density B; n ∈ [500, 6226], density C; n ∈ [500, 577], density D) to evaluate the patch classifier performance by accuracy, area under the receiver operating characteristics curve (AUC), recall, precision, F1 score, false negative rate, and false positive rate. To evaluate hidden bias across subgroups, we assessed model performance by race – White, Black, or Other; age groups – <50, 50-60, 60-70, and >70 years; BI-RADS tissue density – A, B, C, or D; pathology – never biopsied, cancer, or benign; and imaging features – mass, asymmetry, architectural distortion, or calcification.

**Statistical Analysis**

*Univariate analysis with AUC comparisons:* Overall subgroup AUC, false negative rate, and false positive rate was compared using Student's *t* test with Bonferroni Correction on the bootstrapped distributions [58, 59]. For each subgroup of a feature, the overall subgroup false negative rate and false positive rate was compared to the control group of that feature. Control groups selected for race, age group, tissue density, and pathological outcome were White, <50 years, BI-RADS density A, never biopsied, respectively. Statistical significance was defined as a *p*-value <0.05.

*Multivariate logistic regression:* We also evaluated the false negative risk for abnormal patches and false positive risk in normal patches using univariate and multivariate logistic regression models to evaluate the impact of various demographic and imaging/clinical subgroups. This allowed us to control for potential confounding effects between features and could help to explain the underlying contribution of each feature to type I and type II errors of model failure individually. False negative predictions were investigated on abnormal patches by race, age, tissue density, pathological outcome, and image findings. False positive predictions were evaluated on normal patches by race, age group, and tissue density as these features are applicable to the whole image, including normal tissue patches. The odds ratio (OR) for false positive and false negative predictions was calculated between each feature and a selected control group model, and the subsequent risk ratio (RR) was calculated by OR and the non-exposed (correct prediction result, in this case, true positives within all positive patches and true negative within all negative patches) prevalence of the target feature.

## Results

**Patient Characteristics**

Mean patient age was 59.02 ± 11.87 (standard deviation) years. There were 21,273 (40.6%) White patients, 22,321 (42.5%) Black patients, and 8,850 (16.9%) patients with other races. 16,387 (31.2%) patients were <50 years, 14,775 (28.2%) patients were 50-60 years, 12,490 (23.8%) patients were 60-70 years, and 8,792 (16.8%) patients were >70 years. Tissue density distribution for BI-RADS A, B, C, and D was 6,050 (11.5%), 18,544 (35.4%), 23,218 (44.3%), and 4,632 (8.8%), respectively. Distribution of imaging features and pathologic outcomes are summarized in Table 2. Of note, counts by imaging feature could not be determined in the training and validation set. This is because, in this dataset, lesion characteristics (i.e., images descriptors and pathologic outcomes) are assigned on a *per image* basis, not *per patch*; therefore, in images with multiple lesions, it was not possible to determine which lesion characteristics corresponded to which patch. In the test dataset, this was addressed by selecting only exams with one lesion and one ROI such that lesion characteristics could be directly linked to each ROI. Distribution of pathological outcomes by imaging findings are shown in Table 3.

**Univariate Analysis of Patch Classification Performance**

On the test set of 13,390 patches from 5,404 patients, the patch classification model built with ResNet152V2 achieved overall accuracy of 92.6% (95% CI = 92.0–93.2%), AUC of 0.975 (95% CI = 0.972-0.978), recall of 0.927 (95% CI = 0.919-0.935), precision of 0.912 (95% CI = 0.902-0.922), F1 score of 0.919 (95% CI = 0.913-0.925). Detailed performance for training, validation, and test sets are shown in Supplemental Table 1. Detailed classification model performance by breast density across subgroups of race, age, tissue density, pathology outcome, and image findings are described in Table 4 (additional metrics in Supplemental Table 2) and ROC curves by subgroups shown in Fig. 4. Without considering confounding between classes, small but statistically significant differences were found by Student's *t* test between AUCs for nearly all subgroups, including differences in performance across all races, better performance with decreasing age, better performance for breast density B, better performance for biopsied lesions (cancer or benign) as compared to never biopsied lesions, and worse performance for patches with architectural distortion (Fig 5). Classification for cancer patches was perfect in density A and D, however this is likely related to the limited number of samples available. The AUC distribution of all 200 bootstrapped test sets is showing in Fig. 6. Sample patches of True Positive, False Negative, True Negative, and False Positive prediction outcomes by various subgroups are shown in Fig. 7.

**Multivariate Logistic Regression Analysis of Misclassification Risk Ratios**

Comparison of performance metrics by individual subgroups does not control for confounding between subgroups. For example, there is a known relation between cancer risk and increasing breast density. Therefore, to appropriately evaluate model performance across subgroups, we employed a multivariate logistic regression model and calculated the relative risk for false negative and false positive predictions in each subgroup. False negative risk was assessed on all 6,142 abnormal patches in the test set (Table 5) and false positive risk was assessed in all 7,248 normal patches (Table 6).

When using the multivariate model, many features that were previously statistically different from each other no longer showed statistically significant difference. The remaining subgroups with lower relative risk of false negatives were Other race (RR=0.828; p=.050), abnormal patches with benign biopsy results (RR=0.927; p=.011), masses (RR=0.921; p=.010), and asymmetries (RR=0.854; p=.040). The only subgroup with higher relative risk of false negative prediction was architectural distortion (RR=1.037; p<.001).

False positive risk assessed on negative patches could only be considered for features that apply to the whole image – race, age, and breast density. Image features and pathologic outcomes could not be considered since these are patch level features only present in abnormal patches. When considering the relative risk of false positives for negative patches, only density C (RR=1.891; p<.001) and density D (RR=2.486; p<.001) showed higher risk of false positives.

## Discussion

Our study addresses several key factors in the use of deep learning models in screening mammography. First – that standard architectures for abnormality classification underperform in certain subgroups of patients or imaging findings, and that this discrepancy is also likely to be present in models that rely on object detection prior to classification since those models use similar components in their backbone architecture. Second – that subgroup performance evaluation without considering confounding between groups can generate misleading results into model performance analysis. Most subgroup evaluation considers classes like race, age, imaging features, or pathologic outcomes as confounder-free independent variables [20,42,43,60,61,62], when in reality, they are confounded. Race and breast density have known relationships with each other and with the risk of breast cancer and thus should not be considered completely independent. Evaluation of model performance by risk of false positives and false negatives provides insight into where these models may fail during clinical deployment. For example, we found that white patients with higher tissue density may also be subject to an elevated risk of model failure, indicating race is not the primary driver of performance degradation. Another noteworthy finding is that neither tissue density nor race exhibits a significant association with model failure defined by false negatives, but that breast density does significantly affect false positives. This outcome is consistent with our initial hypothesis suggesting that patterns in model failure are (1) different between types of failure, (2) subject to potential confounding effects.

We also noted that patches with benign biopsy results had a lower risk of false negatives than never biopsied, while patches with cancer showed a lower trend than benign biopsy although it did not reach statistical significance. This is intuitive since patches that are suspicious enough to warrant biopsy, and especially those that end up yielding cancer, are 'more abnormal' than patches of tissue that were never biopsied if the radiologist later decided they were benign on a follow-up diagnostic exam. Lastly, patches with architectural distortion had higher risk of false negatives than other findings. We postulate that the reason for this is twofold – first, that architectural distortion is more difficult to identify, even for radiologists; second, that there were fewer samples of AD compared to other imaging subtypes and the model may not have enough examples to accurately identify these abnormalities.

When model failure is defined as false positives, only whole image level demographic and clinical attributes can be assured on negative patches. Only the impacts of different races, age groups, and tissue densities were considered in this analysis accordingly. Our findings suggest that higher density groups, BI-RADS density C and D, have a 1.89 and a 2.48 times higher increased risk of false positives compared to BI-RADS density A patches, respectively. This confirmed the concern and challenges in clinic, that the lesions within high density breast tissues are harder to distinguish. In clinical practice this may lead to more false negatives, and the model results of higher false positives in these regions may have unintended consequences.

The findings should be interpreted considering the study's methodological constraints. First, this research was conducted at a single institution, utilizing medical images collected from multiple hospitals. Consequently, the extent to which our results can be generalized to other healthcare systems requires further investigation, which requires a dataset with similar annotated lesion characteristics at the patch level. Secondly, our study did not conduct a direct failure analysis on an object detection model for breast cancer. Instead, we hypothesize that lesion characteristics associated with reduced performance in patch classification may also lead to a degradation in performance in object detection tasks. The experimental design was adopted due to the absence of a reliable and publicly available object detection model tailored specifically for mammography classification, which challenges direct failure analysis on such a model. Lastly, we only included exams in subgroup analysis for which data was available and grouped imaging features and race into broad categories, which may obscure nuanced differences within these groups.

## Conclusions

The study employed a unique experimental design to investigate the underlying and isolated effects of demographic, clinical, and imaging characteristics on the risk of failure in abnormality patch classification within screening mammograms by addressing potential confounding effects of these factors. The implications of these findings may be applied to support the advancement of computational strategies for classification and object detection, inform the development of more precise and broadly applicable computational tools, and enable the identification of populations that may face disadvantages due to the implementation of such tools if no measures are taken. These are important steps towards improving the performance of fair and interpretable decision-making models in mammography.

**Table 1: Comparison of performance of multiple standard convolutional neural network (CNN) models for binary patch classification of abnormal versus normal patches on mammography. We found ResNet152V2 achieved the highest performance and was therefore used for the remainder of experiments.**

| Model | Accuracy | AUC | Recall | Precision | F1 Score |
|---|---|---|---|---|---|
| VGG16 | 0.794 | 0.881 | 0.794 | 0.766 | 0.780 |
| InceptionV3 | 0.906 | 0.967 | 0.886 | 0.907 | 0.897 |
| ResNet50V2 | 0.918 | 0.968 | 0.912 | 0.909 | 0.910 |
| ResNet152V2 | 0.926 | 0.975 | 0.927 | 0.912 | 0.919 |

AUC = Area Under the receiver operating characteristics Curve

**Table 2: Distribution of Image Patches Used for Model Development.**

| Characteristics | Count (Percentage) | | | |
| --- | --- | --- | --- | --- |
| | Total | Training Set | Validation Set | Test Set |
| All patches | 52,444(100%) | 29,144(100%) | 9,910(100%) | 13,390(100%) |
| **Label** | | | | |
| Positive | 24,166(46.1%) | 13,461(46.2%) | 4,563(46.0%) | 6,142(45.9%) |
| Negative | 28,278(53.9%) | 15,683(53.8%) | 5,347(54.0%) | 7,248(54.1%) |
| **BI-RADS** | | | | |
| 0-additional imaging needed | 38,293(73.0%) | 21,318(73.1%) | 7,212(72.8%) | 9,763(72.9%) |
| 1-negative or 2-benign | 14,151(27.0%) | 7,826(26.9%) | 2,698(27.2%) | 3,627(27.1%) |
| **Race** | | | | |
| African American or Black | 22,321(42.5%) | 12,483(42.8%) | 4,137(41.7%) | 5,701(42.6%) |
| Caucasian or White | 21,273(40.6%) | 11,649(40.0%) | 4,140(41.8%) | 5,484(40.9%) |
| Other | 8,850(16.9%) | 5,012(17.2%) | 1,633(16.5%) | 2,205(16.5%) |
| **Age Group** | | | | |
| <50 | 16,387(31.2%) | 9,184(31.5%) | 3,293(33.2%) | 3,910(29.2%) |
| 50-60 | 14,775(28.2%) | 8,138(27.9%) | 2,811(28.4%) | 3,826(28.6%) |
| 60-70 | 12,490(23.8%) | 6,965(23.9%) | 2,240(22.6%) | 3,285(24.5%) |
| >70 | 8,792(16.8%) | 4,857(16.7%) | 1,566(15.8%) | 2,369(17.7%) |
| **BI-RADS Density** | | | | |
| A-Almost entirely fatty | 6,050(11.5%) | 3,218(11.0%) | 1,166(11.8%) | 1,666(12.4%) |
| B-Scattered areas of fibroglandular density | 18,544(35.4%) | 10,275(35.3%) | 3,348(33.8%) | 4,921(36.8%) |
| C-Heterogeneously dense | 23,218(44.3%) | 12,682(43.5%) | 4,310(43.5%) | 6,226(46.5%) |
| D-Extremely dense | 4,632(8.8%) | 2,969(10.2%) | 1,086(10.9%) | 577(4.3%) |
| **Pathology** | | | | |
| Cancer | 1,147(2.2%) | 739(2.5%) | 233(2.3%) | 175(1.3%) |
| Benign | 3,495(6.7%) | 2,025(7.0%) | 662(6.7%) | 808(6.0%) |
| Never Biopsied | 47,802(91.1%) | 26,380(90.5%) | 9,015(91.0%) | 12,407(92.7%) |
| **Image Finding** | | | | |
| Mass | | | | 1,774(13.2%) |
| Asymmetry | | | | 5,372(40.1%) |
| AD | | | | 579(4.3%) |
| Calcification | | | | 2,502(18.7%) |

AD = Architectural Distortion, BI-RADS = Breast Imaging Reporting and Data System
Due to the structure of the dataset, imaging features prevalence could only be calculated for the test set in which images contained a single ROI. Images in the training and validation dataset may contain multiple ROIs so exact prevalence could not be calculated

**Table 3: Distribution of pathologic outcomes by image findings in the test set.**

| Image Findings | Cancer | Benign | Never Biopsied | Total |
|---|---|---|---|---|
| Mass | 8 (0.45%) | 75 (4.22%) | 1,693 (95.33%) | 1,776 (100%) |
| Asymmetry | 97 (3.86%) | 390 (15.54%) | 2,023 (80.60%) | 2,510 (100%) |
| AD | 19 (3.28%) | 50 (8.64%) | 510 (88.08%) | 579 (100%) |
| Calcification | 82 (1.53%) | 265 (4.93%) | 5,025 (93.54%) | 5,372 (100%) |

AD = Architectural Distortion

**Table 4: Subgroup classification performance stratified by tissue density on the test set**

| Subgroups | Metrics | Density A | Density B | Density C | Density D | Overall |
|---|---|---|---|---|---|---|
| **Overall** | AUC: | 0.977±0.013 | 0.982±0.004 | 0.966±0.005 | 0.953±0.021 | **0.975±0.003** |
| | Recall: | 0.924±0.040 | 0.942±0.012 | 0.920±0.012 | 0.899±0.046 | **0.927±0.008** |
| | Precision: | 0.825±0.069 | 0.936±0.014 | 0.906±0.013 | 0.882±0.047 | **0.912±0.010** |
| **Race** | | | | | | |
| White | AUC: | 0.984±0.011 | 0.978±0.008 | 0.962±0.011 | 0.955±0.006 | 0.972±0.005 |
| | Recall: | 0.920±0.048 | 0.936±0.023 | 0.907±0.026 | 0.882±0.013 | 0.918±0.013 |
| | Precision: | 0.787±0.068 | 0.933±0.020 | 0.895±0.024 | 0.870±0.014 | 0.902±0.016 |
| Black | AUC: | 0.973±0.010 | 0.983±0.005 | 0.967±0.011 | 0.949±0.008 | 0.976±0.005 |
| | Recall: | 0.924±0.031 | 0.941±0.019 | 0.924±0.020 | 0.900±0.016 | 0.931±0.012 |
| | Precision: | 0.846±0.039 | 0.938±0.019 | 0.907±0.022 | 0.874±0.019 | 0.914±0.015 |
| Other | AUC: | 0.988±0.015 | 0.988±0.007 | 0.972±0.011 | 0.949±0.009 | 0.978±0.007 |
| | Recall: | 0.953±0.077 | 0.953±0.026 | 0.930±0.025 | 0.923±0.017 | 0.938±0.020 |
| | Precision: | 0.838±0.142 | 0.940±0.030 | 0.924±0.023 | 0.910±0.016 | 0.926±0.020 |
| **Age Group** | | | | | | |
| <50 | AUC: | 0.984±0.011 | 0.977±0.011 | 0.966±0.010 | 0.951±0.006 | 0.970±0.007 |
| | Recall: | 0.921±0.066 | 0.939±0.027 | 0.919±0.021 | 0.869±0.013 | 0.922±0.014 |
| | Precision: | 0.847±0.086 | 0.940±0.027 | 0.915±0.019 | 0.909±0.013 | 0.919±0.016 |
| 50-60 | AUC: | 0.971±0.024 | 0.986±0.006 | 0.968±0.010 | 0.972±0.005 | 0.977±0.005 |
| | Recall: | 0.906±0.065 | 0.944±0.023 | 0.918±0.023 | 0.974±0.010 | 0.930±0.017 |
| | Precision: | 0.780±0.072 | 0.952±0.017 | 0.905±0.021 | 0.853±0.019 | 0.915±0.018 |
| 60-70 | AUC: | 0.985±0.007 | 0.980±0.007 | 0.965±0.016 | 0.937±0.015 | 0.976±0.006 |
| | Recall: | 0.936±0.036 | 0.934±0.022 | 0.927±0.035 | 0.859±0.027 | 0.932±0.017 |
| | Precision: | 0.880±0.041 | 0.931±0.023 | 0.906±0.035 | 0.911±0.029 | 0.916±0.022 |
| >70 | AUC: | 0.966±0.022 | 0.983±0.008 | 0.962±0.017 | 0.943±0.018 | 0.975±0.008 |
| | Recall: | 0.926±0.054 | 0.949±0.028 | 0.901±0.056 | 0.921±0.035 | 0.928±0.023 |
| | Precision: | 0.780±0.073 | 0.916±0.036 | 0.875±0.050 | 0.766±0.043 | 0.882±0.031 |
| **Pathology** | | | | | | |
| Cancer | AUC: | 1.000[†] | 0.975±0.050 | 0.981±0.032 | 1.000[†] | 0.980±0.023 |
| | Recall: | 1.000[†] | 0.947±0.077 | 0.911±0.126 | 1.000[†] | 0.938±0.063 |
| | Precision: | 1.000[†] | 0.949±0.068 | 0.957±0.097 | 1.000[†] | 0.957±0.052 |
| Benign | AUC: | 0.999±0.004 | 0.981±0.015 | 0.969±0.024 | 0.994±0.004 | 0.977±0.010 |
| | Recall: | 1.000[†] | 0.949±0.039 | 0.949±0.042 | 0.966±0.024 | 0.955±0.022 |
| | Precision: | 0.972±0.045 | 0.955±0.030 | 0.926±0.049 | 0.965±0.023 | 0.943±0.027 |
| Never Biopsied | AUC: | 0.975±0.008 | 0.982±0.004 | 0.966±0.006 | 0.948±0.005 | 0.974±0.003 |
| | Recall: | 0.913±0.025 | 0.940±0.014 | 0.916±0.013 | 0.888±0.009 | 0.925±0.008 |
| | Precision: | 0.809±0.035 | 0.934±0.013 | 0.904±0.014 | 0.870±0.010 | 0.908±0.010 |
| **Image Findings** | | | | | | |
| Mass | AUC: | 0.996±0.003 | 0.984±0.013 | 0.973±0.013 | 0.936±0.014 | 0.980±0.008 |
| | Recall: | 0.979±0.033 | 0.957±0.031 | 0.934±0.033 | 0.952±0.016 | 0.949±0.022 |
| | Precision: | 0.820±0.073 | 0.932±0.036 | 0.889±0.044 | 0.852±0.028 | 0.896±0.029 |
| Calcification | AUC: | 0.987±0.010 | 0.979±0.010 | 0.969±0.013 | 0.967±0.006 | 0.974±0.008 |
| | Recall: | 0.965±0.041 | 0.942±0.026 | 0.933±0.027 | 0.954±0.012 | 0.939±0.018 |
| | Precision: | 0.829±0.076 | 0.926±0.034 | 0.903±0.033 | 0.875±0.016 | 0.904±0.022 |
| AD | AUC: | 0.873±0.222 | 0.947±0.033 | 0.902±0.056 | 0.895±0.031 | 0.914±0.034 |
| | Recall: | 0.595±0.343 | 0.804±0.072 | 0.831±0.066 | 0.718±0.038 | 0.810±0.040 |
| | Precision: | 1.000[†] | 0.985±0.026 | 0.918±0.057 | 0.897±0.035 | 0.939±0.027 |
| Asymmetry | AUC: | 0.983±0.009 | 0.988±0.004 | 0.966±0.009 | 0.973±0.006 | 0.977±0.005 |
| | Recall: | 0.911±0.031 | 0.955±0.014 | 0.924±0.016 | 0.929±0.013 | 0.937±0.010 |
| | Precision: | 0.982±0.016 | 0.963±0.013 | 0.919±0.018 | 0.936±0.012 | 0.942±0.011 |
| **Total Count** | | 1,666 (12.4%) | 4,921 (36.8%) | 6,226 (46.5%) | 577 (4.3%) | 13,390 (100%) |

AD = Architectural Distortion, AUC = Area Under the receiver operating characteristics Curve
The overall and subgroup AUC, recall, and precision averaged over 200 bootstrapped samples ± 95% confidence interval, randomly sized ranging in [500,1666] for patches with BI-RADS density A, [500,4921] for patches with BI-RADS density B, [500,6226] for patches with BI-RADS density C, and [500,577] for patches with BI-RADS density D. Pairwise statistical significance difference of AUC was given in Fig. 5 across subgroups, and statistical significance difference of false negative rate of subgroup versus control groups (White, <50y/o, density A, Never Biopsied, and no such image finding) of abnormal patches was shown in Table 5; false positive rate statistical significance of normal patches was shown in Table 6
[†]: Cases within subgroup were all correctly classified, no confidence intervals applicable

**Table 5: Evaluation of model performance using multivariate logistic regression to assess risk ratio of false negative predictions by subgroups as compared to versus univariate evaluation**

| Variables | OR | RR | Univariate *p*-value | Multivariate *p*-value | Number of Patches | Control Group |
|---|---|---|---|---|---|---|
| Black | 0.880 | 0.922 | <0.001* | 0.248 | 2,630 | White |
| Other* | 0.749 | 0.828 | <0.001* | 0.050* | 1,160 | White |
| 50-60y/o | 0.881 | 0.918 | <0.001* | 0.315 | 1,798 | <50y/o |
| 60-70y/o | 0.823 | 0.875 | <0.001* | 0.163 | 1,434 | <50y/o |
| >70y/o | 0.89 | 0.924 | <0.001* | 0.482 | 840 | <50y/o |
| BI-RADS density B | 1.132 | 1.060 | <0.001* | 0.079 | 311 | BI-RADS density A |
| BI-RADS density C | 0.752 | 0.862 | 0.490 | 0.590 | 2,327 | BI-RADS density A |
| BI-RADS density D | 1.239 | 1.103 | 0.015* | 0.756 | 321 | BI-RADS density A |
| Benign * | 0.567 | 0.927 | <0.001* | 0.011* | 499 | Never Biopsied |
| Cancer | 0.778 | 0.971 | <0.001* | 0.533 | 118 | Never Biopsied |
| Mass * | 0.596 | 0.921 | <0.001* | 0.010* | 761 | No Mass |
| Asymmetry * | 0.751 | 0.854 | <0.001* | 0.040* | 3,127 | No Asymmetry |
| AD * | 2.575 | 1.037 | 0.575 | <0.001* | 413 | No AD |
| Calcification | 0.744 | 0.934 | <0.001* | 0.075 | 1,248 | No Calcification |

AD = Architectural Distortion, BI-RADS = Breast Imaging Reporting and Data System, OR = Odds Ratio, RR = Risk Ratio
Univariate two sample Student's *t* test was conducted to compare difference of false negative rate of bootstrap performance in subgroups versus control groups, as a reference. Demographic and clinical/imaging features were evaluated by multivariate logistic regression models for descriptive analysis to control the potential confounding effect between characteristics. Total number of patches inspected n = 6142
*: Statistically significant, p≤.05

Table 6: **Evaluation of model performance using multivariate logistic regression to assess risk ratio of false positive predictions by subgroups as compared to versus univariate evaluation**

| Variables | OR | RR | Univariate *p*-value | Multivariate *p*-value | Number of Patches | Control Group |
|---|---|---|---|---|---|---|
| Black | 1.107 | 1.058 | 0.153 | 0.306 | 3,071 | White |
| Other | 0.982 | 0.990 | <0.001* | 0.893 | 1,045 | White |
| 50-60y/o | 0.914 | 0.934 | <0.001* | 0.445 | 2,028 | <50y/o |
| 60-70y/o | 0.870 | 0.899 | <0.001* | 0.279 | 1,851 | <50y/o |
| >70y/o | 0.920 | 0.938 | <0.001* | 0.540 | 1,529 | <50y/o |
| BI-RADS density B | 1.328 | 1.249 | <0.001* | 0.071 | 2,594 | BI-RADS density A |
| BI-RADS density C * | 2.406 | 1.891 | <0.001* | <0.001* | 3,043 | BI-RADS density A |
| BI-RADS density D * | 3.863 | 2.486 | 0.015* | <0.001* | 256 | BI-RADS density A |

BI-RADS = Breast Imaging Reporting and Data System, OR = Odds Ratio, RR = Risk Ratio
Univariate two sample Student's *t* test was conducted to compare difference of false positive rate of bootstrap performance in subgroups versus control groups, as a reference. Demographic and clinical features were evaluated by multivariate logistic regression models for descriptive analysis to control the potential confounding effect between characteristics. Variables analyzed are in whole image level, while other lesion level imaging features were not available on negative patches, thus pathological outcomes and image findings were not applicable to a false positive analysis. Total number of patches inspected n = 7248
*: Statistically significant, p≤.05

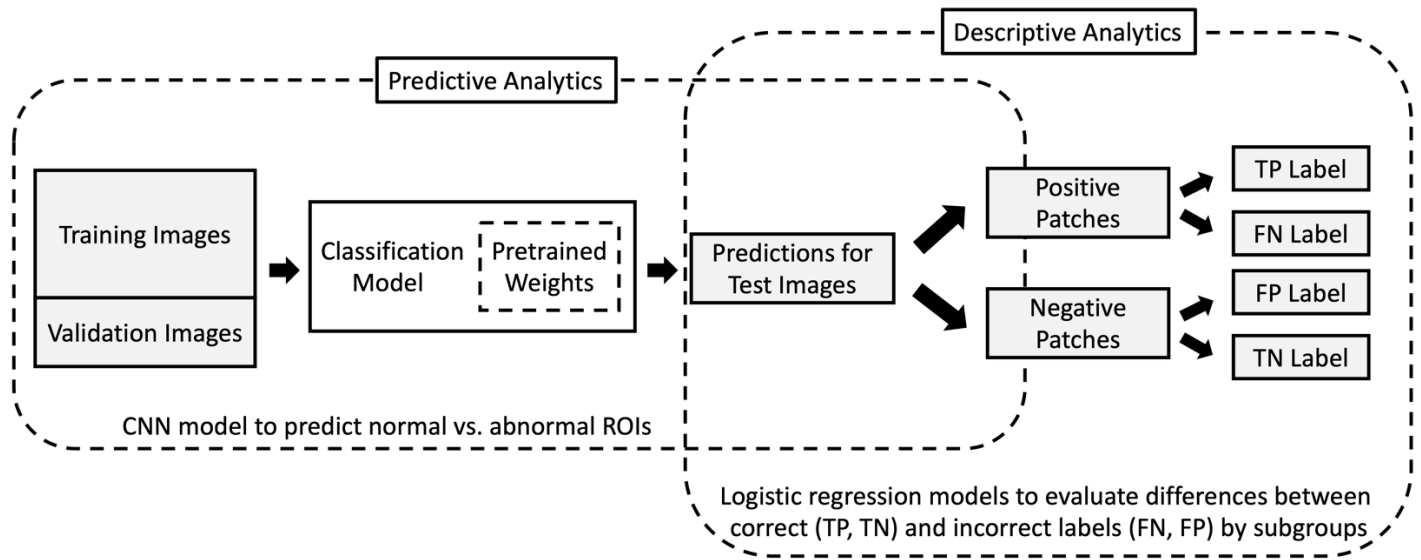

**Fig. 1** Experimental overview with two main components. The component of predictive analytics involves training a deep learning model to distinguish between normal and abnormal patches on mammography. The component of descriptive analytics involves evaluation of subgroup model performance using multivariate logistic regression models to control for confounding between subgroups. TP = True Positive, TN = True Negative, FP = False Positive, FN = False Negative

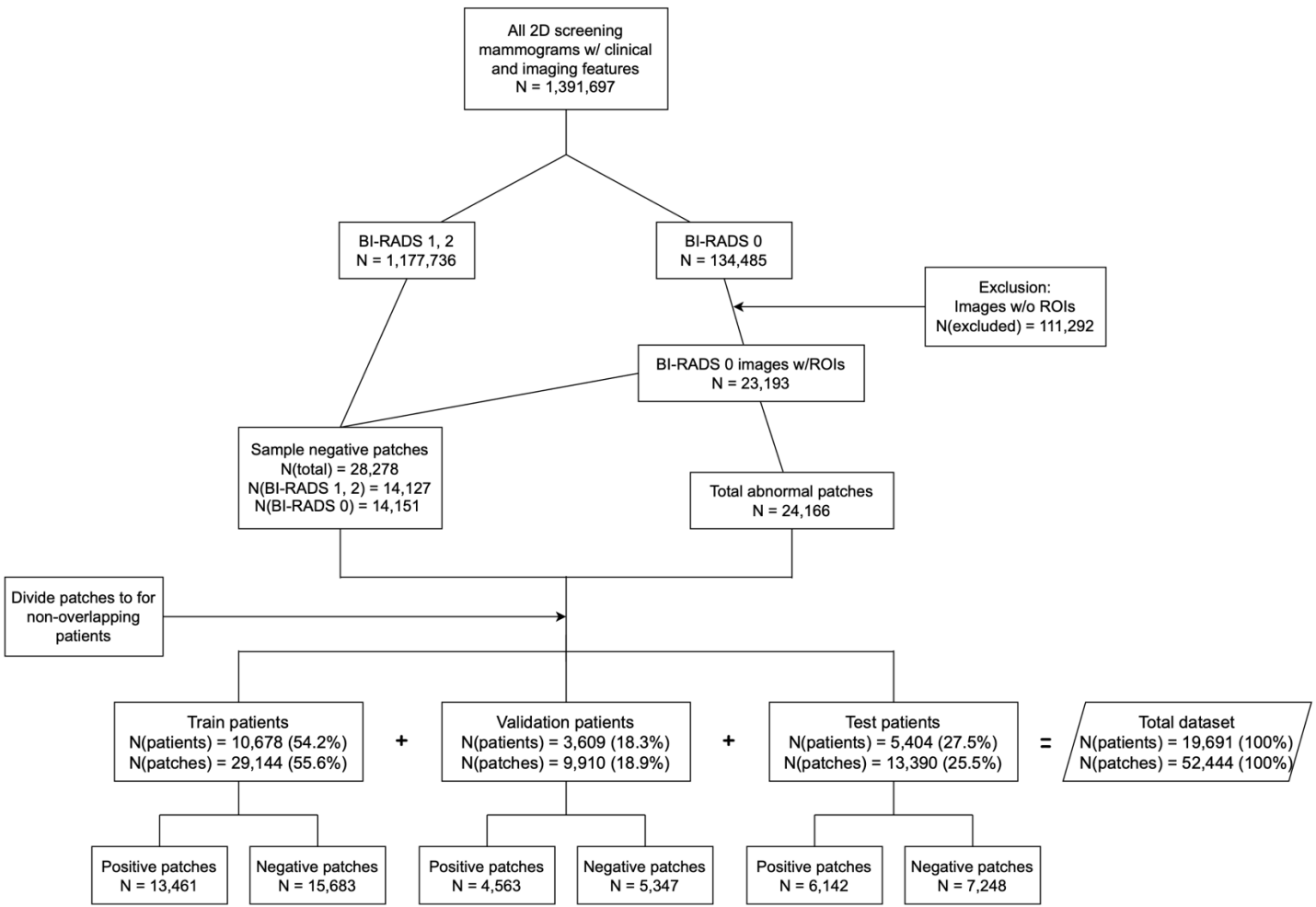

**Fig. 2** Flow chart of dataset creation. Positive patches are extracted from BI-RADS 0 images using annotations created by the radiologist at time of original interpretation. Negative patches are randomly selected from BI-RADS 1 and 2 images and areas outside of the abnormal patches on BI-RADS 0 images. Similar numbers of negative and positive patches were created. Training, validation, and test sets were separated by patient to avoid data leakage. BI-RADS = Breast Imaging Reporting and Data System

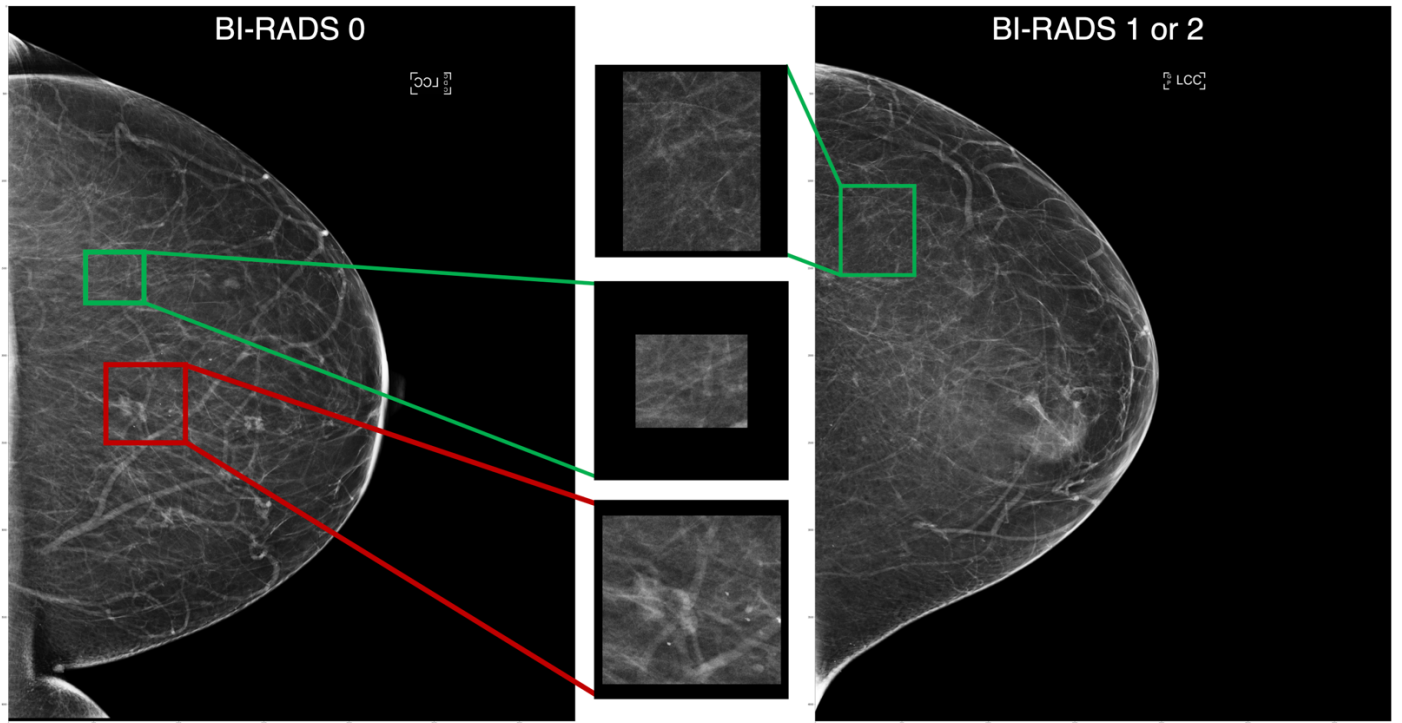

**Fig. 3** Example of positive and negative patch generation. The patch highlighted in the red box represents an abnormality detected by the radiologist on a BI-RADS 0 exam. The patches highlighted in green represent randomly selected other patches from BI-RADS 0 images (left) and negative images (right) within breast tissue areas. The patches were cropped, centered, and padded with 0 pixels to be 512×512 pixels. Patches larger than 512×512 pixels were downsampled and then padded. BI-RADS = Breast Imaging Reporting and Data System

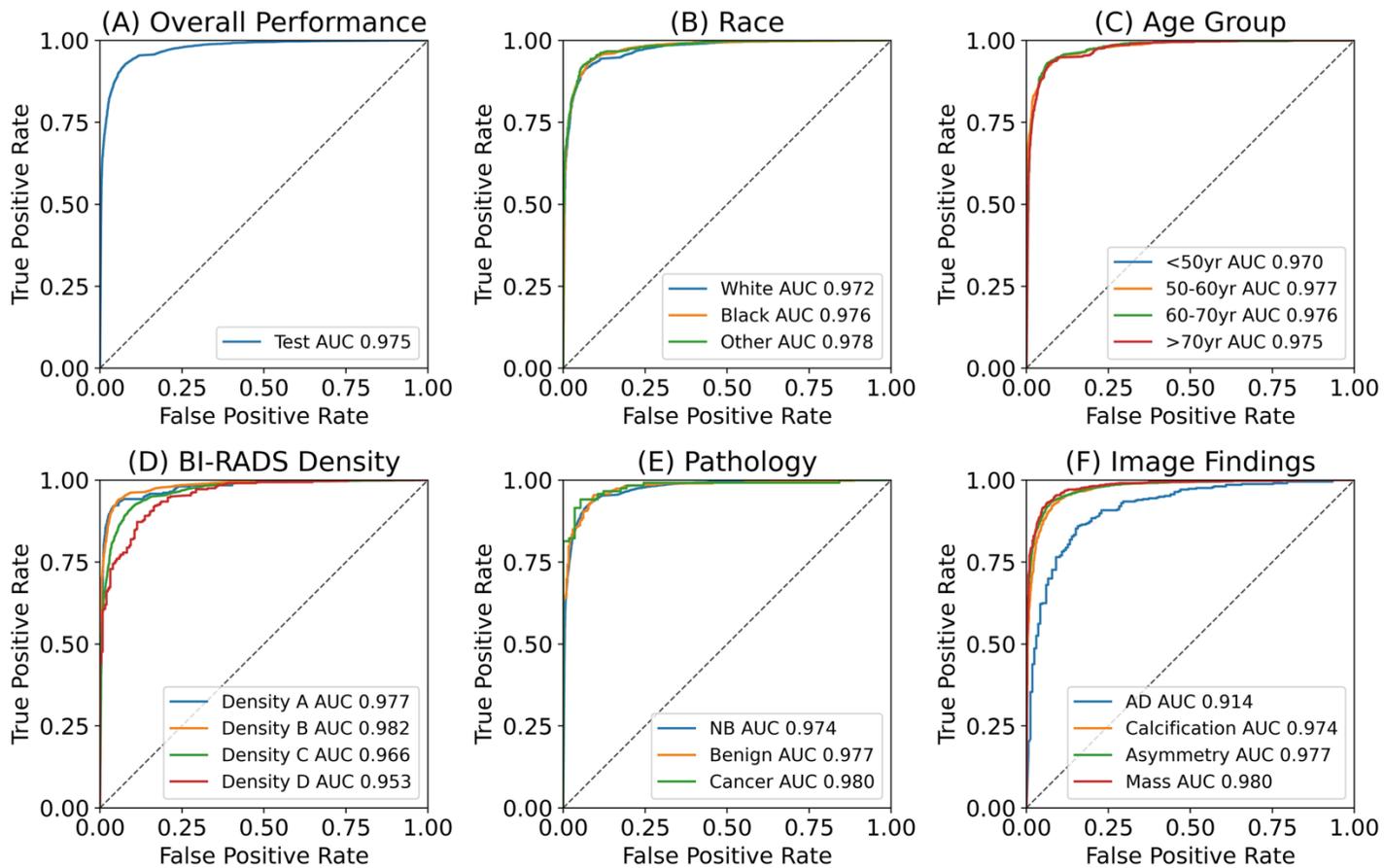

**Fig. 4** Receiver operating characteristic curves of the whole test set **(a)**, and by subgroups – race **(b)**, age group **(c)**, BI-RADS density **(d)**, pathology **(e)**, and image findings **(f)**. AUC = Area Under the receiver operating characteristics Curve, BI-RADS = Breast Imaging Reporting and Data System, NB = Never Biopsied

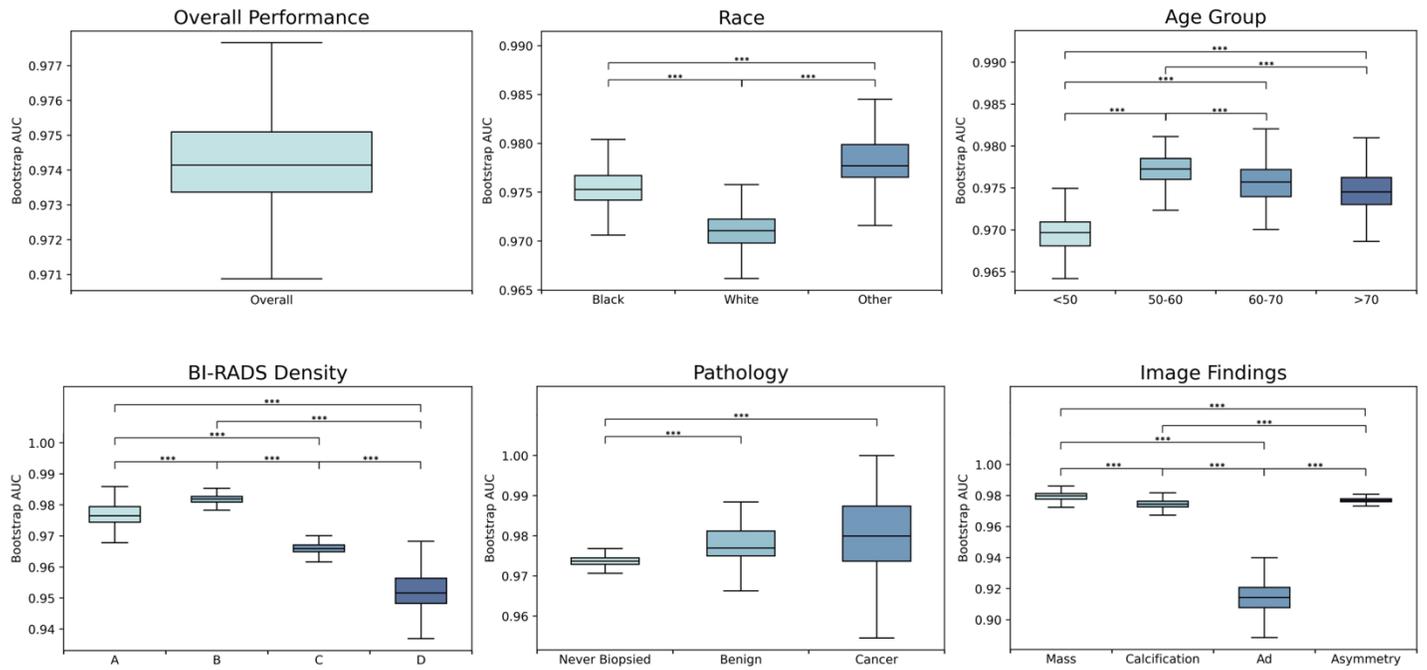

**Fig. 5** Boxplots for 200 bootstrapped AUC overall and by characteristics. The significance of difference in mean were calculated pairwise within subgroups in each characteristic, followed by Bonferroni correction on the p-value. Nearly all subgroups showed small but statistically different differences in performance, with the exception of age groups 60-70 and >70 y/r, and between biopsied benign and cancerous lesions.

AD = Architectural Distortion, AUC = Area Under the receiver operating characteristics Curve, BI-RADS = Breast Imaging Reporting and Data System

Statistical significance level: *** represents p<0.001

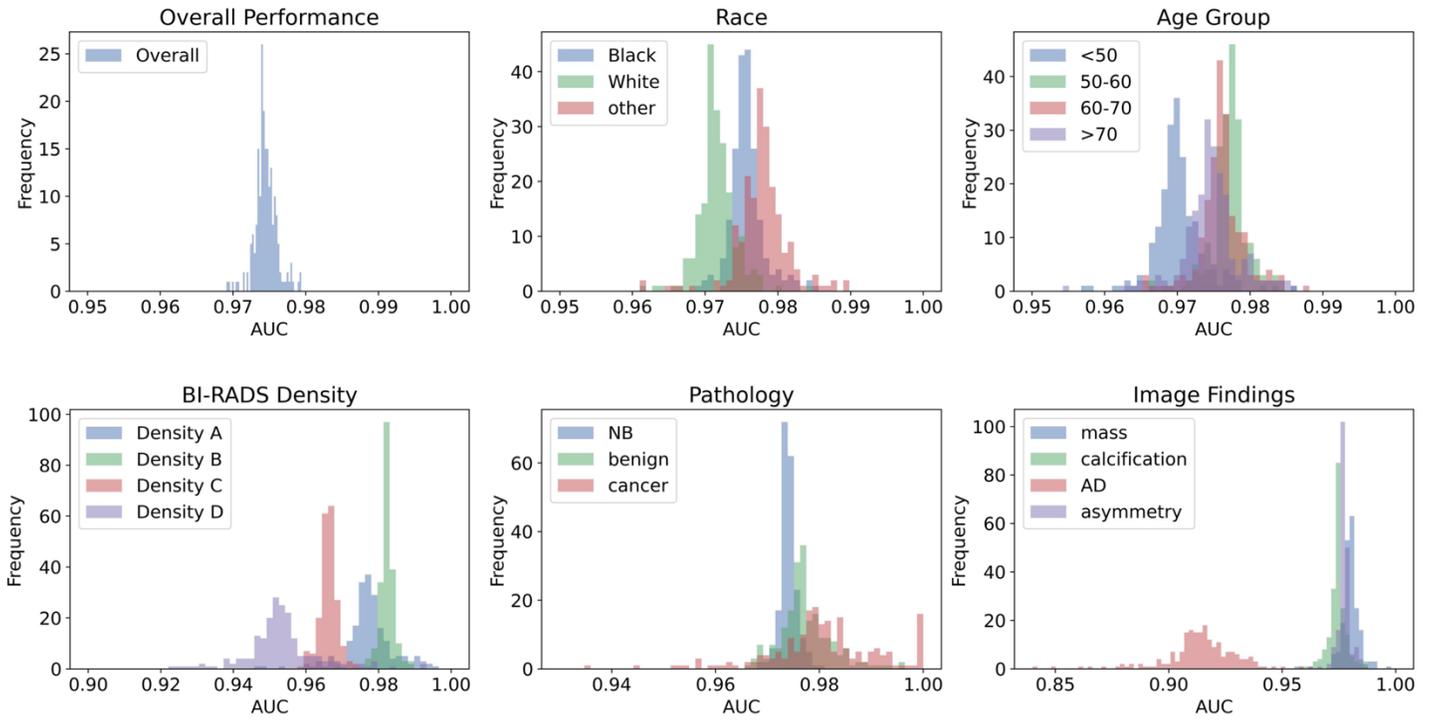

**Fig. 6** Histogram demonstrating distribution of AUCs of the test set with 200 bootstraps showing the separation between subgroups within each category. AD = Architectural Distortion, AUC = Area Under the receiver operating characteristics Curve, BI-RADS = Breast Imaging Reporting and Data System, NB = Never Biopsied

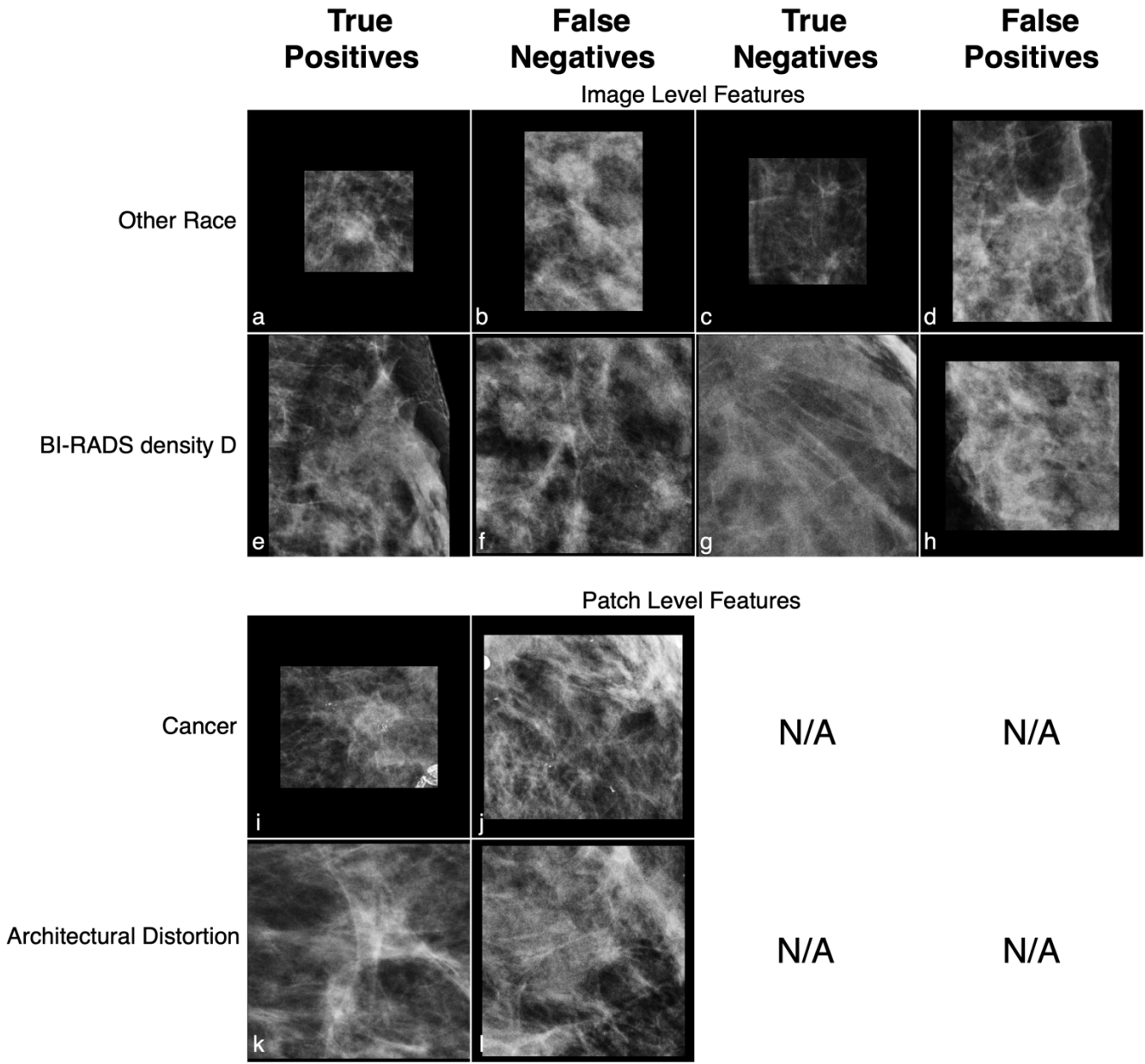

**Fig. 7** Example of True Positive, False Negative, True Negative, and False Positive patches of various patches. Each row is selected to demonstrate a single feature while holding other features constant. **(a)-(d)** are patches selected from patients of Other race, age less than 50 years old, BI-RADS density C, and never biopsied; **(e)-(h)** are patches of patients with density D, race White, age less than 50 years old, and never biopsied; **(i)** and **(j)** are patches with cancer, White, age greater than 70 years old, density B, with calcifications; **(k)** and **(l)** are patches with architectural distortion, White, age less than 50 years old, density C, never biopsied. BI-RADS = Breast Imaging Reporting and Data System

**Supplemental Table 1: ResNet152V2 classification model overall performance and 95% confidence interval in test set**

| Group | Accuracy | AUC | Recall | Precision | F1 Score |
|---|---|---|---|---|---|
| **Train** | 0.984 | 0.998 | 0.984 | 0.983 | 0.983 |
| **Validation** | 0.926 | 0.967 | 0.915 | 0.925 | 0.920 |
| **Test** | 0.926±0.006 | 0.975±0.003 | 0.927±0.008 | 0.912±0.010 | 0.919±0.006 |

AUC = Area Under the receiver operating characteristics Curve

The test set performance averaged over 200 bootstrapped samples ± 95% confidence interval, randomly sized ranging in [1000,13390]

**Supplemental Table 2: Subgroup classification performance stratified by tissue density**

| Subgroups | Metrics Description | Density A | Density B | Density C | Density D | Overall |
|---|---|---|---|---|---|---|
| **Overall** | Accuracy: | 0.950±0.016 | 0.942±0.009 | 0.910±0.008 | 0.877±0.036 | **0.926±0.006** |
| | AUC: | 0.977±0.013 | 0.982±0.004 | 0.966±0.005 | 0.953±0.021 | **0.975±0.003** |
| | Recall: | 0.924±0.040 | 0.942±0.012 | 0.920±0.012 | 0.899±0.046 | **0.927±0.008** |
| | Precision: | 0.825±0.069 | 0.936±0.014 | 0.906±0.013 | 0.882±0.047 | **0.912±0.010** |
| | F1 Score: | 0.871±0.044 | 0.939±0.009 | 0.913±0.009 | 0.891±0.034 | **0.919±0.006** |
| **Race** | | | | | | |
| White | Accuracy: | 0.952±0.016 | 0.943±0.013 | 0.905±0.017 | 0.866±0.010 | 0.922±0.009 |
| | AUC: | 0.984±0.011 | 0.978±0.008 | 0.962±0.011 | 0.955±0.006 | 0.972±0.005 |
| | Recall: | 0.920±0.048 | 0.936±0.023 | 0.907±0.026 | 0.882±0.013 | 0.918±0.013 |
| | Precision: | 0.787±0.068 | 0.933±0.020 | 0.895±0.024 | 0.870±0.014 | 0.902±0.016 |
| | F1 Score: | 0.848±0.049 | 0.934±0.016 | 0.901±0.019 | 0.876±0.010 | 0.910±0.011 |
| Black | Accuracy: | 0.948±0.011 | 0.941±0.013 | 0.910±0.016 | 0.862±0.015 | 0.927±0.009 |
| | AUC: | 0.973±0.010 | 0.983±0.005 | 0.967±0.011 | 0.949±0.008 | 0.976±0.005 |
| | Recall: | 0.924±0.031 | 0.941±0.019 | 0.924±0.020 | 0.900±0.016 | 0.931±0.012 |
| | Precision: | 0.846±0.039 | 0.938±0.019 | 0.907±0.022 | 0.874±0.019 | 0.914±0.015 |
| | F1 Score: | 0.883±0.025 | 0.940±0.014 | 0.915±0.016 | 0.887±0.013 | 0.922±0.010 |
| Other | Accuracy: | 0.961±0.029 | 0.944±0.020 | 0.918±0.017 | 0.909±0.013 | 0.928±0.015 |
| | AUC: | 0.988±0.015 | 0.988±0.007 | 0.972±0.011 | 0.949±0.009 | 0.978±0.007 |
| | Recall: | 0.953±0.077 | 0.953±0.026 | 0.930±0.025 | 0.923±0.017 | 0.938±0.020 |
| | Precision: | 0.838±0.142 | 0.940±0.030 | 0.924±0.023 | 0.910±0.016 | 0.926±0.020 |
| | F1 Score: | 0.891±0.094 | 0.946±0.020 | 0.927±0.017 | 0.916±0.011 | 0.932±0.014 |
| **Age Group** | | | | | | |
| <50 | Accuracy: | 0.955±0.022 | 0.937±0.019 | 0.906±0.016 | 0.871±0.010 | 0.916±0.012 |
| | AUC: | 0.984±0.011 | 0.977±0.011 | 0.966±0.010 | 0.951±0.006 | 0.970±0.007 |
| | Recall: | 0.921±0.066 | 0.939±0.027 | 0.919±0.021 | 0.869±0.013 | 0.922±0.014 |
| | Precision: | 0.847±0.086 | 0.940±0.027 | 0.915±0.019 | 0.909±0.013 | 0.919±0.016 |
| | F1 Score: | 0.882±0.057 | 0.939±0.019 | 0.917±0.015 | 0.888±0.009 | 0.920±0.011 |
| 50-60 | Accuracy: | 0.943±0.016 | 0.947±0.015 | 0.910±0.017 | 0.893±0.014 | 0.926±0.011 |
| | AUC: | 0.971±0.024 | 0.986±0.006 | 0.968±0.010 | 0.972±0.005 | 0.977±0.005 |
| | Recall: | 0.906±0.065 | 0.944±0.023 | 0.918±0.023 | 0.974±0.010 | 0.930±0.017 |
| | Precision: | 0.780±0.072 | 0.952±0.017 | 0.905±0.021 | 0.853±0.019 | 0.915±0.018 |
| | F1 Score: | 0.837±0.052 | 0.948±0.014 | 0.912±0.017 | 0.909±0.012 | 0.922±0.012 |
| 60-70 | Accuracy: | 0.957±0.013 | 0.936±0.014 | 0.918±0.024 | 0.888±0.019 | 0.933±0.013 |
| | AUC: | 0.985±0.007 | 0.980±0.007 | 0.965±0.016 | 0.937±0.015 | 0.976±0.006 |
| | Recall: | 0.936±0.036 | 0.934±0.022 | 0.927±0.035 | 0.859±0.027 | 0.932±0.017 |
| | Precision: | 0.880±0.041 | 0.931±0.023 | 0.906±0.035 | 0.911±0.029 | 0.916±0.022 |
| | F1 Score: | 0.907±0.030 | 0.933±0.017 | 0.917±0.027 | 0.884±0.021 | 0.924±0.016 |
| >70 | Accuracy: | 0.947±0.018 | 0.948±0.017 | 0.907±0.032 | 0.834±0.029 | 0.930±0.014 |
| | AUC: | 0.966±0.022 | 0.983±0.008 | 0.962±0.017 | 0.943±0.018 | 0.975±0.008 |
| | Recall: | 0.926±0.054 | 0.949±0.028 | 0.901±0.056 | 0.921±0.035 | 0.928±0.023 |
| | Precision: | 0.780±0.073 | 0.916±0.036 | 0.875±0.050 | 0.766±0.043 | 0.882±0.031 |
| | F1 Score: | 0.846±0.051 | 0.932±0.024 | 0.888±0.039 | 0.837±0.030 | 0.904±0.021 |
| **Pathology** | | | | | | |
| Cancer | Accuracy: | 1.000† | 0.933±0.070 | 0.914±0.090 | 1.000† | 0.930±0.049 |
| | AUC: | 1.000† | 0.975±0.050 | 0.981±0.032 | 1.000† | 0.980±0.023 |
| | Recall: | 1.000† | 0.947±0.077 | 0.911±0.126 | 1.000† | 0.938±0.063 |
| | Precision: | 1.000† | 0.949±0.068 | 0.957±0.097 | 1.000† | 0.957±0.052 |
| | F1 Score: | 1.000† | 0.948±0.050 | 0.932±0.089 | 1.000† | 0.947±0.042 |
| Benign | Accuracy: | 0.983±0.028 | 0.938±0.032 | 0.923±0.039 | 0.958±0.020 | 0.937±0.019 |
|  | AUC: | 0.999±0.004 | 0.981±0.015 | 0.969±0.024 | 0.994±0.004 | 0.977±0.010 |
|  | Recall: | 1.000† | 0.949±0.039 | 0.949±0.042 | 0.966±0.024 | 0.955±0.022 |
|  | Precision: | 0.972±0.045 | 0.955±0.030 | 0.926±0.049 | 0.965±0.023 | 0.943±0.027 |
|  | F1 Score: | 0.986±0.023 | 0.952±0.026 | 0.937±0.034 | 0.965±0.017 | 0.949±0.016 |
| NB | Accuracy: | 0.949±0.008 | 0.942±0.009 | 0.909±0.010 | 0.866±0.008 | 0.925±0.006 |
|  | AUC: | 0.975±0.008 | 0.982±0.004 | 0.966±0.006 | 0.948±0.005 | 0.974±0.003 |
|  | Recall: | 0.913±0.025 | 0.940±0.014 | 0.916±0.013 | 0.888±0.009 | 0.925±0.008 |
|  | Precision: | 0.809±0.035 | 0.934±0.013 | 0.904±0.014 | 0.870±0.010 | 0.908±0.010 |
|  | F1 Score: | 0.858±0.023 | 0.937±0.009 | 0.910±0.011 | 0.879±0.007 | 0.916±0.007 |
| **Image Findings** | | | | | | |
| Mass | Accuracy: | 0.960±0.016 | 0.949±0.023 | 0.912±0.029 | 0.883±0.020 | 0.931±0.017 |
|  | AUC: | 0.996±0.003 | 0.984±0.013 | 0.973±0.013 | 0.936±0.014 | 0.980±0.008 |
|  | Recall: | 0.979±0.033 | 0.957±0.031 | 0.934±0.033 | 0.952±0.016 | 0.949±0.022 |
|  | Precision: | 0.820±0.073 | 0.932±0.036 | 0.889±0.044 | 0.852±0.028 | 0.896±0.029 |
|  | F1 Score: | 0.892±0.046 | 0.944±0.025 | 0.911±0.031 | 0.899±0.018 | 0.922±0.020 |
| Calcification | Accuracy: | 0.947±0.024 | 0.932±0.024 | 0.911±0.023 | 0.889±0.013 | 0.920±0.016 |
|  | AUC: | 0.987±0.010 | 0.979±0.010 | 0.969±0.013 | 0.967±0.006 | 0.974±0.008 |
|  | Recall: | 0.965±0.041 | 0.942±0.026 | 0.933±0.027 | 0.954±0.012 | 0.939±0.018 |
|  | Precision: | 0.829±0.076 | 0.926±0.034 | 0.903±0.033 | 0.875±0.016 | 0.904±0.022 |
|  | F1 Score: | 0.891±0.049 | 0.934±0.024 | 0.918±0.022 | 0.913±0.010 | 0.921±0.016 |
| AD | Accuracy: | 0.710±0.272 | 0.848±0.056 | 0.829±0.052 | 0.732±0.033 | 0.826±0.032 |
|  | AUC: | 0.873±0.222 | 0.947±0.033 | 0.902±0.056 | 0.895±0.031 | 0.914±0.034 |
|  | Recall: | 0.595±0.343 | 0.804±0.072 | 0.831±0.066 | 0.718±0.038 | 0.810±0.040 |
|  | Precision: | 1.000† | 0.985±0.026 | 0.918±0.057 | 0.897±0.035 | 0.939±0.027 |
|  | F1 Score: | 0.748±0.218 | 0.885±0.045 | 0.872±0.045 | 0.797±0.029 | 0.869±0.026 |
| Asymmetry | Accuracy: | 0.944±0.019 | 0.952±0.011 | 0.908±0.014 | 0.919±0.010 | 0.930±0.009 |
|  | AUC: | 0.983±0.009 | 0.988±0.004 | 0.966±0.009 | 0.973±0.006 | 0.977±0.005 |
|  | Recall: | 0.911±0.031 | 0.955±0.014 | 0.924±0.016 | 0.929±0.013 | 0.937±0.010 |
|  | Precision: | 0.982±0.016 | 0.963±0.013 | 0.919±0.018 | 0.936±0.012 | 0.942±0.011 |
|  | F1 Score: | 0.945±0.018 | 0.959±0.009 | 0.921±0.013 | 0.932±0.009 | 0.939±0.008 |
| **Total Count** | | 1,666 (12.4%) | 4,921 (36.8%) | 6,226 (46.5%) | 577 (4.3%) | 13,390 (100%) |

AD = Architectural Distortion, AUC = Area Under the receiver operating characteristics Curve, BI-RADS = Breast Imaging Reporting and Data System, FNR = False Negative Rate

The overall and subgroup performance averaged over 200 bootstrapped samples ± 95% confidence interval, randomly sized ranging in [500,1666] for patches with BI-RADS density A, [500,4921] for patches with BI-RADS density B, [500,6226] for patches with BI-RADS density C, and [500,577] for patches with BI-RADS density D; for all tissue densities overall, bootstrap samples ranged in [1000,13390]

†: Cases within subgroup were all correctly classified, no confidence intervals applicable